\documentclass[superscriptaddress,twocolumn,amsmath,amssymb,footinbib]{revtex4-1}
\usepackage[english]{babel}
\usepackage[pdftex]{graphicx}
\usepackage{gensymb}
\begin{document}

\newcommand{\physicsDept}{Department of Physics, Laboratory of Atomic and Solid State Physics, Cornell University, Ithaca, NY 14853, USA}
\newcommand{\mseDept}{Department of Materials Science and Engineering, Cornell University, Ithaca, NY 14853, USA}
\newcommand{\aepDept}{School of Applied and Engineering Physics, Cornell University, Ithaca, New York 14853, USA}
\newcommand{\kavli}{Kavli Institute at Cornell for Nanoscale Science, Ithaca, NY 14853, USA}
\newcommand{\Uchicago}{Pritzker School of Molecular Engineering, The University of Chicago, Chicago, IL 60637, USA}
\newcommand{\ALS}{Advanced Light Source, E.O. Lawrence Berkeley National Laboratory, Berkeley, CA 94720, USA}
\newcommand{\RO}{RuO$_2$ }
\newcommand{\ROnospace}{RuO$_2$ }
\newcommand{\FeSeSTO}{FeSe/SrTiO$_3$ }
\newcommand{\FeSeSTOns}{FeSe/SrTiO$_3$}
\newcommand{\TO}{TiO$_2$ }
\newcommand{\TOnospace}{TiO$_2$}
\newcommand{\TC}{$T_{c0}$ }
\newcommand{\TCc}{$T_{c}$ }
\newcommand{\Tc}{$T_{c}$ }
\newcommand{\TCnospace}{$T_c$}
\newcommand{\ang}{\text{\normalfont\AA}}
\newcommand{\dpar}{$d_{||}$ }
\newcommand{\dparnospace}{$d_{||}$}
\newcommand{\dxzyz}{$(d_{\text{xz}}$, $d_{\text{yz}})$ }
\newcommand{\dxzyznospace}{$(d_{\text{xz}}$, $d_{\text{yz}})$}
\newcommand{\Tgap}{$T_{\Delta}$ }
\newcommand{\Tgapns}{$T_{\Delta}$}
\newcommand{\Tonset}{$T_{onset}$ }
\newcommand{\Tonsetns}{$T_{onset}$}
\newcommand{\Tzero}{$T_{0}$ }
\newcommand{\Tzerons}{$T_{0}$}
\newcommand{\Tbkt}{$T_{BKT}$ }
\newcommand{\Tbktns}{$T_{BKT}$}
\newcommand{\Tstar}{$T^{\ast}$ }
\newcommand{\Tstarns}{$T^{\ast}$}
\newcommand{\insitu}{\emph{in situ} }
\newcommand{\exsitu}{\emph{ex situ} }
\newcommand{\Rsheet}{$R_{s}$ }
\newcommand{\Rsheetns}{$R_{s}$}

\title{Interfacial Electron-Phonon Coupling Constants Extracted from Intrinsic Replica Bands in Monolayer \FeSeSTOns}

\author{Brendan D. Faeth}\altaffiliation{Corresponding author: bdf53@cornell.edu} 
\affiliation{\physicsDept}
\author{Saien Xie}
\affiliation{\physicsDept}
\author{Shuolong Yang}\altaffiliation{Current Address: \Uchicago} 
\affiliation{\physicsDept}
\affiliation{\kavli}
\affiliation{\mseDept}
\author{Jason K. Kawasaki}\altaffiliation{Current Address: Department of Materials Science and Engineering, University of Wisconsin, Madison, Wisconsin 53706, USA.} 
\affiliation{\physicsDept}
\affiliation{\kavli}
\author{Jocienne N. Nelson}
\affiliation{\physicsDept}
\author{Shuyuan Zhang}
\affiliation{\physicsDept}
\author{Pramita Mishra}\altaffiliation{Current Address: Department of Physics, Indian Institute of Science, Bangalore 560012, India}
\affiliation{\physicsDept}
\author{Chen Li}
\affiliation{\physicsDept}
\author{Christopher Jozwiak}
\affiliation{\ALS}
\author{Aaron Bostwick}
\affiliation{\ALS}
\author{Eli Rotenberg}
\affiliation{\ALS}
\author{Darrell G. Schlom}
\affiliation{\mseDept}
\affiliation{\kavli}
\author{Kyle M. Shen}\altaffiliation{Corresponding author: kmshen@cornell.edu} 
\affiliation{\physicsDept}
\affiliation{\kavli}


\begin{abstract}
The observation of replica bands by angle-resolved photoemission spectroscopy has ignited interest in the study of electron-phonon coupling at low carrier densities, particularly in monolayer \FeSeSTOns, where the appearance of replica bands has motivated theoretical work suggesting that the interfacial coupling of electrons in the FeSe layer to optical phonons in the SrTiO$_3$ substrate might contribute to the enhanced superconducting pairing temperature. Alternatively, it has also been recently proposed that such replica bands might instead originate from extrinsic final state losses associated with the photoemission process. Here, we perform a quantitative examination of replica bands in monolayer \FeSeSTOns, where we are able to conclusively demonstrate that the replica bands are indeed signatures of intrinsic electron-boson coupling, and not associated with final state effects. A detailed analysis of the energy splittings between the higher-order replicas, as well as other self-energy effects, allow us to determine that the interfacial electron-phonon coupling in the system corresponds to a value of $\lambda = 0.19 \pm 0.02$. 

\end{abstract}
\maketitle

One of the most powerful attributes of angle-resolved photoemission spectroscopy (ARPES) is its ability to reveal many-body interactions through its lineshape, owing to its close relationship to the single-particle spectral function $A(\mathbf{k}, \omega)$. ARPES has revealed the presence of strong electron-boson coupling in a variety of quantum materials, including high-temperature cuprate superconductors~\cite{lanzara_evidence_2001, zhou_universal_2003}, colossal magnetoresistive manganites~\cite{mannella_nodal_2005}, and titanates~\cite{Wang2016}. At high carrier densities, electron-boson coupling is manifested as an abrupt kink in the quasiparticle dispersion occurring at the boson energy. At low carrier densities, where screening is weaker and the Fermi energy, $E_F$, can be comparable to the relevant phonon frequency, $\Omega_0$, the electron-phonon coupling can give rise to polaronic quasiparticles and the presence of satellite ``replica bands'', which occur at near-integer multiples of $\Omega_0$. Such features have been recently reported in a variety of systems, including at the surface of SrTiO$_3$~\cite{Wang2016,Chen2015,Zhang2017}, anatase TiO$_2$~\cite{PhysRevLett.110.196403,Verdi2017}, and most notably in monolayer FeSe films grown on SrTiO$_3$~\cite{Lee2014,Song2019,Shi2017,Yang2019}, where it has been argued that the interfacial coupling of electrons in the FeSe monolayer to optical phonons in the SrTiO$_3$ substrate could potentially be responsible for its enhanced superconducting properties~\cite{Lee2014}. Furthermore, it may be possible to extract more extensive quantitative information about the nature of interactions through a detailed analysis of the replica bands, including their intensities and energy separations. On the other hand, it has also been recently suggested that these replica bands observed by ARPES are not signatures of intrinsic electron-phonon interactions, but rather could arise from extrinsic electron energy losses in the photoemission process, whereby ejected photoelectrons lose energy to surface phonons~\cite{Sawatzky2018}. Such extrinsic ``final-state effects'' produced by photoelectron energy loss processes would appear very similar to intrinsic satellites. Further complicating the situation, such loss features should be prevalent in systems where the carrier density is low and screening effects are weak, precisely where polaronic quasiparticles and intrinsic replica bands in ARPES would also be expected. 

Therefore, a detailed investigation of photoemission replica bands is imperative to distinguish whether they are indeed intrinsic features in the spectral function of quantum materials versus extrinsic final-state loss effects. Given the importance of ARPES as the premier tool for investigating electronic many-body interactions, this is critical not only for the understanding of ARPES as a technique, but also for the general study of quantum materials and their many-body interactions.  To achieve this, we investigate MBE-grown, single layer \FeSeSTO thin films, where such replica bands have been observed, but also where it has been suggested as potentially arising from extrinsic loss effects from Fuchs-Kliewer phonons in the SrTiO$_3$ substrate~\cite{Fuchs1965}. By employing a wide range of photon energies, we are able to conclusively determine that the replica bands indeed arise from intrinsic electron-phonon coupling between the FeSe film and SrTiO$_3$ substrate, and not from extrinsic losses. Furthermore, a quantitative analysis of the spectra in comparison to prior theoretical calculations also allows us to determine the coupling constant, $\lambda = 0.19 \pm 0.02$, by extracting the blue shift of the first satellite feature, as well as accurately measuring the ratio of the main band to replica bands~\cite{PhysRevB.100.241101}. This work not only demonstrates that replica bands in \FeSeSTO arise from intrinsic electron-phonon coupling, but also suggests a generalized approach for evaluating whether features in the photoemission spectrum indeed arise from intrinsic many-body interactions in other quantum materials. Furthermore, this work introduces a new experimental procedure for extracting quantitative electron-phonon coupling constants in polaronic systems from observed replica bands, for instance, allowing us to estimate the likely enhancement of $T_c$ in monolayer \FeSeSTO due to coupling to the interfacial SrTiO$_3$ phonons. 

Monolayer FeSe films were grown by molecular beam epitaxy (MBE) on undoped SrTiO$_3$ substrates and measured immediately by \emph{in situ} ARPES (He-I photons, $h\nu$ = 21.2 eV) as well as \emph{in situ} resistivity measurements (Fig.~S1). Having verified their quality and superconducting properties, samples were then capped with $\approx$~100~nm amorphous Se for transport to the Advanced Light Source MAESTRO beamline (7.0.2) in a sealed, inert environment. Films were then decapped at the endstation at 420$^\circ$ C, in a vacuum better than \smash{5 $\times$ 10$^{-10}$} Torr immediately prior to ARPES measurements. ARPES measurements were then performed at 15 K at photon energies ranging between 21-75 eV, with a total energy resolution of 10-20 meV (depending on the incident photon energy).

\begin{figure}
\begin{center}
\includegraphics[width=3.4 in]{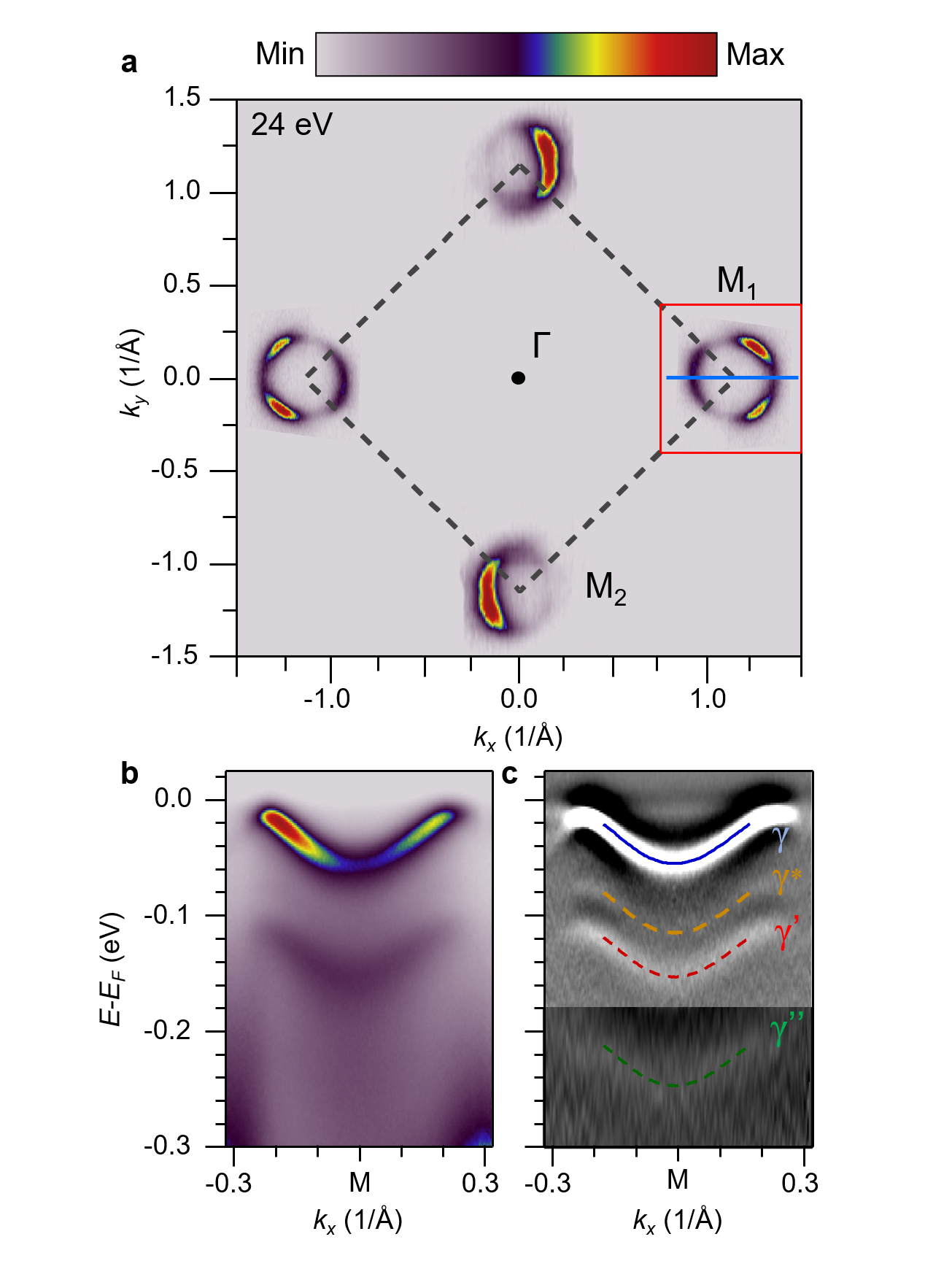}
\end{center}
\caption{Fermi surface and replica band topology in single-layer \FeSeSTOns. (\textbf{a})~Fermi surface map of single-layer \FeSeSTO taken with $p$-polarized light at $h\nu$ = 24eV.  (\textbf{b})~High-statistics spectra along the cut shown at $M_1$ (blue).  (\textbf{c})~Second-derivative of the spectra in (b).  An additional 60~meV replica (labelled $\gamma^{*}$ in the figure) is clearly visible, as well as a second-order replica (labelled $\gamma''$) after saturating the color scale over the higher binding energy region.}  
\label{fig:Fig1}
\end{figure}

\begin{figure}
\begin{center}
\includegraphics[width=3.4in]{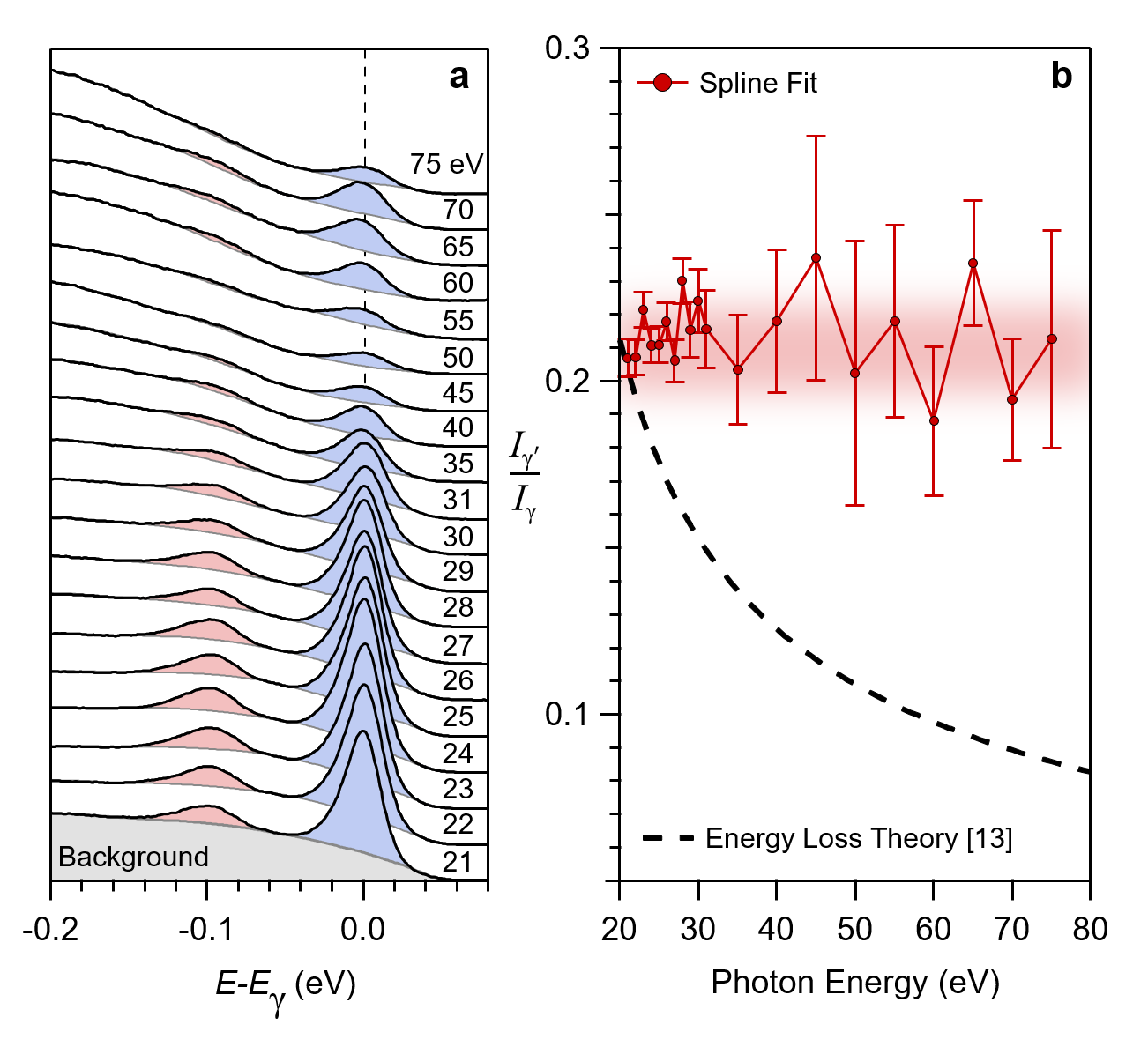}
\end{center}
\caption{Photon energy dependence of the replica band intensity. (\textbf{a})~Integrated EDC's collected from $h\nu$ = 21 to 75~eV.  Grey lines indicate the spline background, while blue and red shaded regions indicate the integrated signal of the $\gamma$ and $\gamma'$ bands, respectively.  (\textbf{b})~Relative intensity of $\gamma'$ to $\gamma$ as a function of incident photon energy (red markers) compared to the theoretical prediction for a photoelectron loss effect from Ref. \cite{Sawatzky2018} (black dashed line).}  
\label{fig:Fig4}
\end{figure}

\begin{figure}
\begin{center}
\includegraphics[width=8.47cm]{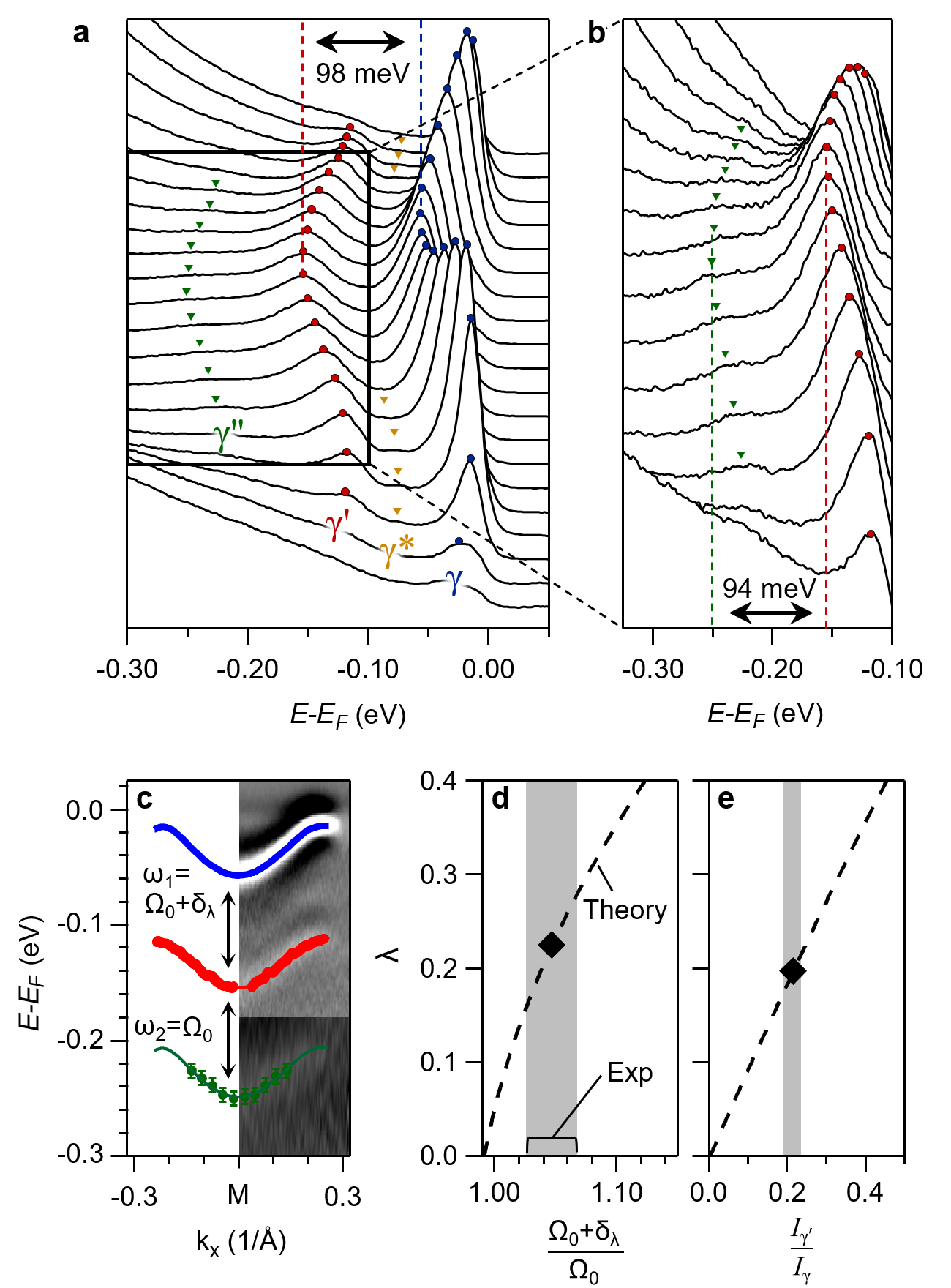}
\end{center}
\caption{Observation of second-order replica bands in single-layer \FeSeSTOns. (\textbf{a,b})~EDC's across the spectra at $M$ shown as a waterfall plot.  Blue, red, yellow, and green markers track the main band ($\gamma$), 98~meV replica ($\gamma'$), and 60 meV replica ($\gamma^*$), and 190 meV second order replica ($\gamma''$ respectively.  (\textbf{c}) Band positions based on fits to the EDC peak positions.  (\textbf{d,e})~Determination of the electron-phonon coupling constant $\lambda$ based on the $\gamma'$ blue shift (d) and replica band intensity (e).  Theoretical behavior based on Ref. \cite{PhysRevB.100.241101}.  Grey regions indicate the experimental uncertainty.}  
\label{fig:fig2}
\end{figure}

In Figure~\ref{fig:Fig1} we present ARPES measurements of monolayer \FeSeSTO after decapping, taken at $h\nu$ = 24 eV with $p$-polarized light. Consistent with previous reports, the Fermi surface is comprised of nearly degenerate elliptical electron pockets at $M$ which arise from the glide-mirror symmetry of the iron-selenium plane~\cite{ZX_PRL2016}.  Due to matrix element effects, the electron pocket at M$_1$ appears with lobes of increased intensity in a three-fold pattern about the pocket.  In Figure~\ref{fig:Fig1}(b,c), we show raw and second-derivative spectra taken along the $\Gamma$-$M$ direction as indicated by the blue line in Fig. 1(a). The main band closest to $E_F$, which we denote as $\gamma$, exhibits a well-defined gap and back-bending at 15 K, indicative of the expected superconductivity. The first replica band, ($\gamma'$, red) is visible in the raw ARPES spectra 98 meV below the main band.  Two additional, weaker features are also visible in the second-derivative spectra shown in Fig. 1(c), including a faint replica at $\approx$60 meV (denoted as ($\gamma^*$, shown in orange), and also an additional replica band separated by 192 meV from the main band ($\gamma''$, denoted in green). Their characteristic energies associate them with two distinct Fuchs-Kliewer (F-K) phonons of the SrTiO$_3$ substrate, which arise from out-of-plane vibration modes of the oppositely charged Ti and O ions~\cite{PhysRevB.38.5699}; $\gamma^*$ corresponds to FK$_2$, and $\gamma'$ and $\gamma''$ are the first and second-order satellites from FK$_1$~\cite{zhang2016}. 


If the replica bands indeed arise from extrinsic final-state energy losses, as suggested in Ref. \cite{Sawatzky2018}, then it is predicted that the ratio of the intensity of the first replica band, $\gamma'$, relative to the intensity of the main band, $\gamma$, $I_{\gamma'}/I_{\gamma}$, should depend strongly on the kinetic energy and direction of the outgoing photoelectron. Conversely, if the replica bands arise from electron-phonon coupling in the initial state, they should be intrinsic features of the single-particle spectral function and hence, the intensity ratio $I_{\gamma'}/I_{\gamma}$ should be insensitive to the photoelectron kinetic energy. In  Figure 2(a), we plot energy distribution curves (EDCs) around $M$ between 21 eV $<$ $h\nu$ $<$ 75 eV (corresponding to photoelectron kinetic energies between $\approx$ 17 to 71 eV). To improve statistics, the EDCs have been generated by integrating over the entire band, offsetting each individual EDC by the peak position of the main band $\epsilon_k$ (Fig. S2), then fitted to a smooth spline background with the integrated weight of the main band peak ($I_{\gamma}$) and replica band peak ($I_{\gamma'}$) shown after subtraction in Figure S3. In Figure 2(b), we plot the ratio of $I_{\gamma'}/I_{\gamma}$ as a function of photon energy, which is clearly independent of photon energy, with an extracted value of $I_{\gamma'}/I_{\gamma} = 0.21 \pm 0.02$, together with a comparison of the prediction for the extrinsic photoelectron energy loss scenario, where a $\approx$60\% reduction in $I_{\gamma'}/I_{\gamma}$ would have been expected. This is despite the fact that the overall absolute intensity of both $I_{\gamma}$ and $I_{\gamma'}$ drop by a factor of 10 in going to higher photon energies (hence the larger error bars for $h\nu$ $>$ 40 eV), thus definitively ruling out extrinsic photoelectron loss effects as the origin of the replica band in \FeSeSTOns. We have confirmed that this behavior is robust against details of the fitting procedure, for example, whether a single EDC at $M$ is used as opposed to band-averaged spectra, or whether a Shirley background is used in place of a spline fit. 

Performing this extensive photon energy dependence study provides the opportunity to extract more detailed information about the electron-phonon coupling constant than had previously been possible. For one, this analysis allows us to reliably determine the absolute intensity of $I_{\gamma'}/I_{\gamma}$ = 0.21 $\pm$ 0.02. Furthermore, we are now able to identify an optimal photon energy, $h\nu = 24$ eV with \textit{p}-polarization where the intensity of features is strongest, so as to enable a more detailed, quantitative lineshape analysis of the spectral function and replica bands. In Figure 3(a,b), we show a series of EDCs around $M$ at 24 eV, with the band positions for $\gamma$, $\gamma^*$, $\gamma'$, and $\gamma''$ indicated by markers. This data allows us to accurately determine the separation between $\gamma$ and $\gamma'$ as $\omega_1 = 98 \pm 1$ meV, as well as the separation between the first and second replicas, $\gamma'$ and $\gamma''$, $\omega_2 = 94 \pm 2$ meV, 4 meV less than $\omega_1$. The energy of the FK$_1$ phonon in undoped SrTiO$_3$ substrate has previously been determined to be 94 meV ~\cite{CONARD1993382,Song2019}, although this value is highly doping dependent~\cite{shuyuan2018}. Therefore, the separation between $\gamma$ and $\gamma'$  is blue-shifted by $\delta_\lambda$ = 4 meV, relative to the bare FK$_1$ phonon energy, as shown in Fig. 3(c).  This is in contrast to the separation between $\gamma'$ and $\gamma''$, which closely matches $\Omega_{\text{KF1}}$ to within experimental error ($94 \pm 2$ meV). Such a blue shift of the first-order replica (and corresponding lack of a shift for the second-order replica), has been discussed theoretically in various systems \cite{PhysRevB.100.241101,Verdi2017,PhysRevB.98.094509,Rademaker_2016}, but to our knowledge, this is the first instance where this behavior has been clearly identified experimentally. As has been discussed theoretically, reliably extracting both the blue shift, $\delta_\lambda$, as well as the intensity ratio between the main and first replica bands, $I_{\gamma'}/I_{\gamma}$, allows us to more accurately infer the precise value of the interfacial electron-phonon coupling constant, $\lambda$. This is of particular importance to the \FeSeSTO system, since the possible enhancement of T$_c$ due to coupling to interfacial substrate phonons has been shown to vary strongly as a function of $\lambda$ in certain models~\cite{Rademaker_2016,PhysRevB.100.241101}. Here, we compare our measurements to prior calculations based on solutions of the Migdal-Eliashberg equations of \FeSeSTO in the presence of strong forward-scattering~\cite{PhysRevB.98.094509}. The calculations by Li \emph{et al.} provide theoretical estimates in the limit of small momentum transfer for both $I_{\gamma'}/I_{\gamma}$ and $\delta_\lambda$, which are reproduced as dashed lines in Figures 3(d,e), together with our experimentally determined values for both quantities (shown as shaded bars, which denote our experimental uncertainty). Both $I_{\gamma'}/I_{\gamma} = 0.21 \pm 0.02$, as well as $\delta_\lambda = 4 \pm 2.5$ meV coincide with the theoretical predicted values for $\lambda = 0.19 \pm 0.02$. With this value of $\lambda$=0.19, the work by Li \emph{et al.} would suggest an interfacial enhancement of $\Delta$ of $\approx$ 11 \%. While considerable, this cannot entirely account for gap closing temperature of 60 K compared to bulk electron-doped FeSe compounds (T$_c$ $\approx$ 40 K). In addition, recent combined \textit{in situ} resistivity and ARPES measurements suggest the presence of a pseudogap above 40 K and the possible additional role of the enhanced two-dimensionality of the electronic structure in the monolayer limit~\cite{faeth2020incoherent}.

\begin{figure}
\begin{center}
\includegraphics[width=3.4 in]{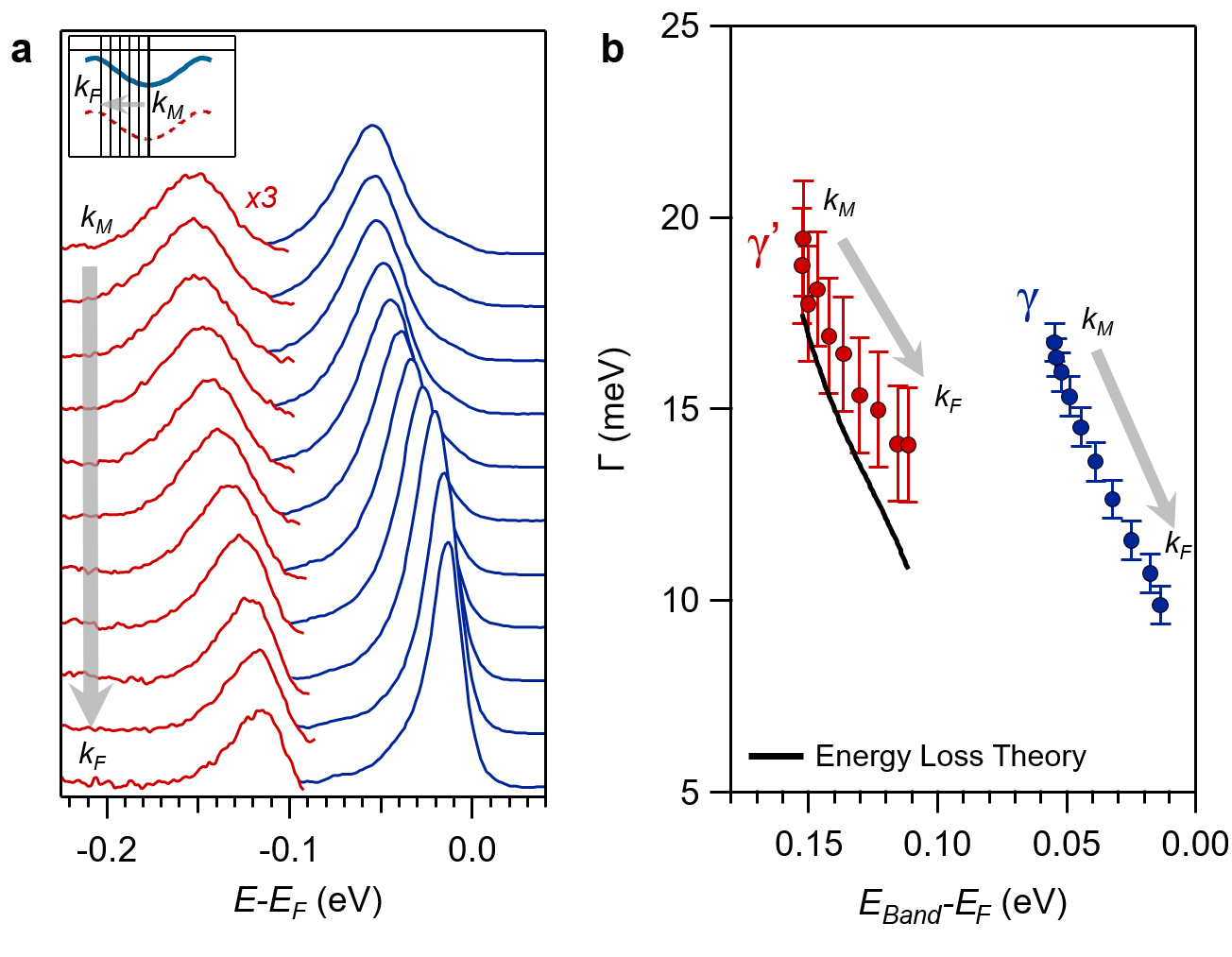}
\end{center}
\caption{Quasiparticle lifetime broadening in the replica band.  (\textbf{a})~Background-subtracted EDC's spanning the electron pocket dispersion, from the band bottom ($k_M$) to $k_F$, for data taken at $h\nu$ = 24 eV.  The $\gamma'$ feature has been multiplied by 3 for visual clarity.  (\textbf{b})~Quasiparticle lifetimes based on spectral function fits for $\gamma$ (blue) and $\gamma'$ (red).  The solid black line indicate the anticipated replica band behavior under the extrinsic photoelectron energy loss scenario.}
\label{fig:Fig3}
\end{figure}

In addition to the blue-shift and intensity ratios, a detailed analysis of the spectral function also reveals evidence for electron-phonon coupling in the lifetime broadening of the first replica band, $\Gamma_{\gamma'}$ relative to that of the main band, $\Gamma_{\gamma}$ (the second replica, $\gamma''$, is too weak to allow a reliable analysis of its lineshape). In Figure 4(a), we show the extracted spectral function after fitting to the backgrounds used in Figure 2 (as before, our conclusions here are independent of the specific background that is employed, Fig. S4). The expected sharpening of the main band peak, $\Gamma_{\gamma}$, as it approaches $E_F$ can be clearly observed in Figure 4(b), where we plot $\Gamma$ as a function of binding energy $E_B$. Likewise, the scattering rate of the first replica band, $\Gamma_{\gamma'}$, exhibits similar behavior, but with the minimum value of $\Gamma_{\gamma'}$ (at $k_F$) approximately equal to the maximum value of $\Gamma_{\gamma}$ (taken at the band bottom, $M$), as would be expected if both features naturally arise from a single, intrinsic spectral function. For comparison, we also plot the simulated linewidth of the first replica, $\Gamma_{\gamma'}$ in the extrinsic photoelectron loss scenario, where the width would correspond to that of the main band $\gamma$ convoluted with the lifetime of the FK$_1$ phonon, $\Gamma_{\text{FK1}}$~\cite{shuyuan2018}. As can be seen in Fig. 4(b), the experimentally determined value of $\Gamma_{\gamma'}$ is substantially larger than would be expected in a photoelectron loss scenario, once again pointing towards its intrinsic character. 

In summary, we have performed an extensive quantitative analysis of replica bands in the ARPES lineshape of single-layer \FeSeSTOns, which allows us to reliably extract both the blue-shift of the first replica band, $\delta_\lambda$, and the intensity ratio $I_{\gamma'}/I_{\gamma}$ between the replica and main bands. A comparison with theoretical calculations in the limit of strong forward-scattering allows us to accurately determine the strength of the coupling between electrons in the FeSe layer and Fuchs-Kliewer phonons in the SrTiO$_3$ substrate as $\lambda = 0.19 \pm 0.02$, suggesting that such interfacial coupling may play an important role in the electronic properties of monolayer \FeSeSTO including its superconductivity, and could also be a promising future approach for investigating the influence of phonons at heterointerfaces. More generally, our approach also provides a clear and unambiguous methodology for distinguishing whether replica bands observed in ARPES spectra of quantum materials with low carrier densities arise from intrinsic electron-phonon coupling or from extrinsic losses associated with the photoemission process.

\section*{Acknowledgments}
This work was supported through the Air Force Office of Scientific Research Grant No. FA9550-15-1-0474, and the National Science Foundation [Platform for the Accelerated Realization, Analysis, and Discovery of Interface Materials (PARADIM)] under Cooperative Agreement No. DMR-1539918, NSF DMR-1709255. This research is funded in part by the Gordon and Betty Moore Foundation's EPiQS Initiative through Grant No. GBMF3850 to Cornell University. B.D.F. and J.N.N. acknowledge support from the NSF Graduate Research Fellowship under Grant No. DGE-1650441.  P.M. acknowledges support from the Indo US Science and Technology Forum (IUSSTF).  This work made use of the Cornell Center for Materials Research (CCMR) Shared Facilities, which are supported through the NSF MRSEC Program (No. DMR-1719875). Substrate preparation was performed in part at the Cornell NanoScale Facility, a member of the National Nanotechnology Coordinated Infrastructure (NNCI), which is supported by the NSF (Grant No. ECCS-1542081). 

\bibliography{Faeth}

\end{document}